\documentstyle[12pt,twoside]{article} 
\begin{document}
\input{epsf} 
\Large
\bf
\begin{center}
      {NUCLEAR ATTENUATION of CHARGED}
\end{center}
\begin{center}
      {MESONS}
\end{center}
\begin{center}	
      IN DEEP INELASTIC SCATTERING
\end{center}

\normalsize
\begin{center}	
      N.Z.Akopov, G.M.Elbakian, L.A.Grigoryan	
\end{center}

\begin{center}	
      Yerevan Physics Institute, Br.Alikhanian 2, 375036 Yerevan, Armenia      	
\end{center}

\begin{center}	
\end{center}

\begin {abstract}

\hspace*{1em}We propose extended version of stationary string model to
describe the nuclear attenuation. This model takes into account flavour 
content of particles 
and allows to include into consideration all hadrons created from string.
The predictions of the model are compared with experimental results
obtained by HERMES collaboration (DESY) on different nuclei (N and Kr).
\end {abstract}

\bf
\Large
\hspace*{1em}1.\hspace*{1em}Introduction\\

\normalsize
\hspace*{1em} The study of the hadrons production in deep inelastic
lepton-nucleus scattering offers the possibility to investigate quark
propagation in dense nuclear matter and the space-time evolution of
hadronization. In particular, the measurements of high energy
hadrons attenuation in nuclear matter is a well-known tool to specify the
parameters of models related to the early stage of particle
production.\\
\hspace*{1em}Hadron production in deep inelastic
scattering (DIS) of leptons on nucleons goes through two stages. On
the first stage 
the virtual photon knocks out the (anti)quark from nucleon, which on the
second stage produce a observable hadron. The strong
interactions theory - perturbative Quantum Chromo Dynamics 
(QCD) at present can not decribe the process of quark hadronization
because of in this case "soft" interactions play a greater role. Therefore, 
experimental
investigation of quark hadronization and building of phenomenological
models describing the most common characteristics of given process is of
importance for development of QCD.\\
\hspace*{1em}Let us consider the simplest possible state, a color-singlet
quark-antiquark system. The lattice QCD studies lend support to a linear
confinement picture, i.e. the energy stored  in the color dipole field
between a charge and an anticharge increases linearly with the separation
between the charges. This is quite different from the behaviour in QED and
is related to the presence of a triple-gluon vertex in QCD. If the tube is
assumed to be uniform along its length, this automatically leads to a
confinement picture with a linearly rising potential. From hadron
spectroscopy the string tension, i.e. the amount of energy per unit
length is deduced to be $\kappa$ $\simeq$ 1 GeV/fm \cite{sjo}. The linear
confinement provide a simple explanation for the existence of linearly
rising Regge trajectories. In this scheme the string tension determined by
the Regge trajectory slope: $\kappa$ =
${(2\pi\alpha^{{\hspace{1mm}^{\prime}}}_{R})}^{-1}$, which is also
$\simeq$ 1 GeV/fm \cite{cas,gur}.\\
\hspace*{1em} Hadronization process takes place at distances of
few fm, right after the deep inelastic interaction with nucleon. As 
this corresponds to the size of the nucleus, it is clear that answers
to these 
questions can be obtained through investigating the processes of
hadron production in DIS of leptons on nuclear targets. The available
experimental results 
were obtained on different nuclear targets with (anti)neutrino 
\cite{ber,bur}, muon \cite{arr,ash}, electron \cite{osb} and positron 
\cite{air} beams. The experiments showed the presence of nuclear 
attenuation effect for charged hadrons with $x_F$ $>$ 0.1 ($x_F$ is 
Feynman's variable), which is essential in the energy 
range of $\sim$ 10 GeV and almost vanish at energies of more than 50 GeV.
It means that the multiplicity of 
hadrons on nuclei is different from the multiplicity on 
deuterium (both per nucleon). The experiments showed also the 
strong dependence of this effect on the energy transfered by leptons to 
nucleus, for electroproduction processes it is energy of virtual 
photon.\\
\hspace*{1em}In the meantime, experimental investigations let to
development of phenomenological models. For description of nuclear
attenuation, mainly the stationary string model (SSM) was used
\cite{bil,czy1}. 
It is supposed that after the DIS of lepton on 
internuclear nucleon, a color string is stretched between the knocked out
quark and the nucleon remnant,  a string which consists of gluons 
with constant tension over it's length. The color field of the string creates
quark-antiquark and diquark-antidiquark pairs, which lead to breaking of 
original string into many short strings. This process results in 
creation of many string-hadrons. The quark which is on the fast
end of string, while passing through the nucleus, can exchange the color
with
one of the quarks of the nucleon, lying within it's trajectory. As a
result, the string breaks on two strings, the energy of the main
string reduces and as a consequence, less fast hadrons will be produced 
in nucleus (per nucleon), than on free nucleon. It is also
presumed, that energy and
momentum of interacting quarks change insignificantly (color
interaction). Some 
authors suppose that the color interaction cross-section is constant
$\sigma^{c}_{q}$ = 
const \cite{gul,czy1}, others believe that it depends on $Q^2$
\cite{kop2}($Q^2$=-$q^2$, where q - is the 4-momentum of the virtual
photon). 
We return to this question later. The model 
supposes that the fast end of the string always contains a quark. It
means that the validity
of the model is limited by the valence region $x_{Bj}$ $>$ 0.2 ($x_{Bj}$
is Bjorken's variable), because in the "sea"
region, it is likely that an antiquark can be on the fast end of the
string, and model does not describe the antiquark interaction. The quark may 
also deviate from it's primary movement direction as a result of
interaction with
nucleon by means of Pomeron or other Reggeon exchange.\\
\hspace*{1em} Let us also briefly consider other models for nuclear
attenuation. In \cite{kop3} authors propose a model of hadronization of
highly virtual quark in nuclear environment by means of gluon bremsstrahlung
and the deceleration of the quark as a result of radiative energy loss. This 
model practically transforms into the SSM at $Q^2$ $<$
3 Ge$V^2$. Recently in \cite{xia} the authors supposed
that in 
deeply inelastic eA collisions in the framework of multiple parton 
scattering, the quark fragmentation functions are modified due to 
higher-twist effects. The hadronization of the quark
takes 
place outside of the nucleus, which is correct in very high energy
ranges. The model has lower limit of application over $Q^2$, which is
not clearly stated by the authors. Also, unrealistic nuclear density 
distribution 
functions for middle and heavy nuclei is used. These questions are important
in sense of application to the HERMES kinematics.\\
\hspace*{1em}Conclusion is that in virtual photon energy region $\nu$ = 
5 - 25 GeV and $<$$Q^2$$>$ equal 2 - 3 Ge$V^2$ which 
are interesting for us in connection with the HERMES experiment  
\cite{air}, most suitable phenomenological model which can be applied for 
quantitative calculations is SSM. Therefore, let us 
outline those points of this model which are improved in present
paper:\\
\hspace*{1em} - description in final state only hadrons summed
over kinds and charges;\\
\hspace*{1em}- the simple description of nucleon structure without "sea"
partons is used. It leads to  the limitation of applicability of the SSM
only for valence region ($x_{Bj}$$>$0.2);\\
\hspace*{1em}- using only part of 
the quark-nucleon cross section connected with color interaction;\\
\hspace*{1em}- the problem connected with description of the $Q^2$
dependence of quark-nucleon and "hadron"-nucleon cross-sections, where 
"hadron" means colorless quark-antiquark system on early stage of it 
creation without "sea".\\
\hspace*{1em} We propose development of SSM, including in
consideration the kind and charge of hadron 
in final state. For this goal we take into account flavours 
and flavour contents of (anti)quarks and diquark from initial
nucleon and quark-antiquark pairs produced in color field of string and also
of final hadron.\\
\hspace*{1em}Second innovation is that in framework of some 
fragmentation scheme we calculate the partial energy $z_h$ = $E_h$/$\nu$ 
(where $E_h$ and $\nu$ are energy of hadron and virtual photon in target
rest frame, respectively) and constituent formation length for any 
hadron produced from string. It allows to calculate the nuclear
attenuation with high accuracy. To simplify the calculations 
we limited ourselves only a three fastest 
hadrons. In other hand because of all calculations in this work were 
performed with the cut on $z_h$ $>$ 0.1-0.2 , then one can take in account 
the limited number of fast hadrons in the string. This is reasonable 
because of  according to  multi-periferical model the parton energy during 
hadronization is divided between the final hadrons in the following 
proportion: $z_{h1}$ $\sim$ 1/2; $z_{h2}$ $\sim$ 1/4; $z_{h3}$ $\sim$ 1/8
and
etc.\\
\hspace*{1em}We considered in final state only charged mesons, because the
mechanism of (anti)protons has some peculiarities, consequently our 
model in present time does not claim to describe the  nuclear
attenuation 
of (anti)protons or $h^+$ and $h^-$. We hope to include them in 
consideration in near future.\\

\bf
\Large
\hspace*{1em}1.\hspace*{1em}Description of developed approach\\

\normalsize
\hspace*{1em}The proposed approach is based on the SSM, 
in which it is  supposed that into the nucleus at the point with 
longitudinal coordinate \textbf{z} and the impact vector \textbf{b} the DIS 
takes place on one of nucleons (proton or neutron). Between the knocked 
out (anti)quark and nucleon remnant the color string is stretched . The 
maximal length of string is L$\simeq$$\nu$/k (\textbf{k} is the string 
tension). Then 
the  breaking of string takes place by means of quark-antiquark and  
diquark-antidiquark 
pair production in color field. The first constituent (anti)quark of 
the hadron with definite kind, charge and partial energy $z_h$ is 
produced on a distance $l_c$ from the point of deep 
inelastic interaction. 
The  $l_c$ is satisfied the condition 
0 $\leq$ $l_c$ $\leq$ $l_{cmax}$, where $l_{cmax}$ is the maximal distance 
on 
which the first constituent of the hadron may create, $l_{cmax}$ = (1 -
$z_h$)$\nu$/k .This relation was derived in 
\cite{kop} based on more general reasons 
than the string model . However this relation could be explained very 
simple in the framework of the string model if we suppose that hadron 
in own rest frame is the string with  $l_0$ = $m_h$/k length, and in the 
system where 
the hadron energy is equal $E_h$, l=$l_0$$E_h$/$m_h$ and $l_c$ = \textbf{L} - 
\textbf{l}, thus the relation  is realized.\\ 
Usually to describe the nuclear attenuation effect the normalized (per 
nucleon) ratio  of hadrons multiplicity produced on nuclear and deuterium 
targets is used. This ratio could be expressed via the 
ratio of corresponding fragmentation functions:\\
\begin{eqnarray}
R_{A}^{h}(z_h)=
\frac{2D_{A}^{h}(z_h)}{AD_{D}^{h}(z_h)}
\end{eqnarray}       
\hspace*{1em}
where $D_{D}^{h}$($z_h$) and $D_{A}^{h}$($z_h$) are the quark
fragmentation
functions in deuterium (assumed same as in vacuum) and in \textbf{A} 
nucleus respectively.\\
\begin{eqnarray}
D_{A}^{h}(z_h) =E(x_{Bj}) 
\int{d^{2}b}\int_{-\infty}^{\infty}{dz\rho(b,z))
\int_{0}^{l_cmax}}
dl_{c}T_{h}(z+l_{c},{\infty})\sum_{i=1}^{n}C_{fi}^{h}
(A,x_{Bj},Q^2)D_{i}^{c}(z_h,l_{c})\\ \nonumber
[F_{i}(z_h)T_{q}(z,z+l_{c})
+ \int_{z}^{z+l_c}
dz^{\hspace{1mm}^{\prime}}\sigma_{q}^{c}(z^{\hspace{1mm}^{\prime}}) 
\rho(b,z^{\hspace{1mm}^{\prime}})F_{i}({z_h}^{\hspace{1mm}^{\prime}})
T_{q}(z^{\hspace{1mm}^{\prime}},z+l_{c})]
\end{eqnarray}

\hspace*{1em}and respectively for deuterium:\\
\begin{eqnarray}
D_{D}^{h}(z_h)=\int_{0}^{l_cmax}
dl_{c}[\sum_{i=1}^{n}C_{fi}^{h}
(2,x_{Bj},Q^2)D_{i}^{c}(z_h,l_{c})F_{i}(z_h)]
\end{eqnarray}

\hspace*{1em}in (2) \footnote{Strictly speaking the expression (2) is 
valid in valence region $x_{Bj}$ $>$ 0.2, since in this region  leading parton 
in string is quark. In "sea" region leading parton may be also antiquark, 
which  color interaction cross section can be different from it for quark 
$\sigma^{c}_{\bar q}$ $\neq$ $\sigma^{c}_{q}$. However for the sake of 
simplicity we do not take into account this difference.} 
and (3) the following notations are used:\\
\begin{eqnarray} 
T_{q(h)}(a,b)= exp(-\int_{a}^{b}dz_{1}
\sigma_{q(h)}(z_{1})\rho(b,z_{1}))
\end{eqnarray}

\hspace*{1em}where:
\hspace*{1em} \textbf{b} - is the impact parameter;\\ 
\hspace*{1em}(\textbf{b},\textbf{z}) -  
is the coordinate of point in which the DIS takes place on one of 
internuclear nucleons;\\ 
\hspace*{1em}(\textbf{b},\textbf{$z^{\hspace{1mm}^{\prime}}$}) - is the 
point at
which the color interaction of the (anti)quark located at the
end 
of string with one of the
internuclear nucleons being on trajectory of movement of knocked out
quark takes place;\\ 
\hspace*{1em}E($x_{Bj}$) - function, which takes into account the 
EMC-effect, the
suppression of nuclear structure function at large values of the $x_{Bj}$
\cite{kop2};\\
\hspace*{1em}$\rho$(r) - is nuclear density function. To parametrize this 
density the shell model \cite{elt} was used  for light nuclei
, and the Woods-Saxon parametrization is used \cite{bil2} for middle and 
heavy nuclei.The normalization
condition is $\int$dr$\rho$(r)=A;\\ 
\hspace*{1em}${z_h}^{\hspace{1mm}^{\prime}}$=$z_h$$\nu$/$\nu$(t),
description of $\nu$(t) will be given after expression (6);\\
\hspace*{1em}\textbf{n} - is the ordinal number of 
hadron produced from the
string. The hadron created at the fast end of string (usually it is called 
the leading hadron) has the ordinal number n = 1, hadron created behind 
it has the ordinal number n =2 and etc. To simplify the 
numerical calculations the limited values
\textbf{n} $\leq$ 3. were used\\ 
\hspace*{1em}$C^{h}_{fi}$(A,$x_{Bj}$,$Q^2$) - are
the functions, which take into account the probability 
of 
hadron's \textbf{h} creation 
 (with given type, charge and ordinal number \textbf{i} in string) from 
quarks and antiquarks being into the  
internuclear
nucleon on which the DIS takes place and q\={q} pairs created in color 
field of the string. The detailed shape 
for this 
function will be presented in Appendix A.\\ 
\hspace*{1em}$D^{c}_{i}$($z_h$,$l_{c}$) - are
constituent formation length $l_c$ distribution functions for hadron
\textbf{i} from string which carry away the partial energy $z_h$.
Detailed shape for these functions is given in Appendix B.\\
\hspace*{1em}$F_i$($z_h$) - are probabilities that i-th hadron produced
from
string carry away the part of the virtual photon energy $z_h$.  
Detailed shape for this function is given in Appendix C.\\
\hspace*{1em}$\sigma_{h}$(z) is the hadron-nucleon total cross section.\\ 
\hspace*{1em}$\sigma_{q}$(z) is the total cross section of quark-nucleon
interaction which could be presented as a sum of two components.\\
\begin{eqnarray}
\sigma_{q}(z) = \sigma^{c}_{q}(z) + \sigma^{0}_{q}(z)
\end{eqnarray}
\hspace*{1em}where $\sigma^{c}_{q}(z)$ is part of the total cross section,
which is connected with color interaction without essential change of 
energy
and momentum of interacting partons. This part of cross section can be
parametrize according to \cite{kop2}:\\
\begin{eqnarray} 
\sigma^{c}_{q}(t) = \frac{C}{Q^{2}(t)} 
\end{eqnarray}
\hspace*{1em} t = $z^{\hspace{1mm}^{\prime}}$ - z;\hspace{5mm}$Q^2$(t) =
$\frac{\nu(t){Q^2}}{\nu+t{Q^2}}$;\hspace{5mm}$\nu$(t) = $\nu$ - kt;
\hspace*{3em}C = 1.32 mbGe$V^2$.\\

\hspace*{1em}Parametrization (6) has bottom limit of validity which
is equal 
$Q^2$ = $Q^{2}_{0}$ = 0.06 Ge$V^2$. At $Q^2$ $<$ $Q^{2}_{0}$  
${\sigma}_{q}^{c}$($Q^2$) = C/(Q$Q_0$) is using. The shape of (6) is
obtained from the idea of 
color transparency, therefore naturally suppose that other part of quark 
nucleon cross section has the the same shape:\\ 
\begin{eqnarray} 
\sigma^{0}_{q}(t) = \frac{C^{\hspace{1mm}^{\prime}}}{Q^{2}(t)}
\end{eqnarray}
\hspace*{1em}where $C^{\hspace{1mm}^{\prime}}$ is the constant generally 
speaking different from \textbf{C}. 
However in order to escape the 
supplementary fit we assume in this work that $C^{\hspace{1mm}^{\prime}}$ 
= C. Really in hadronization process we have deal not with isolate
quark, but with quark which together with antiquark compose the string
or piece of string (hadron). This system is colorless quark-antiquark
dipole. Consequently, instead of (7) we can use the phenomenological
dipole 
cross section. For example simple and convenient parametrization has
been suggested in \cite{kop4}. Easy to see that for the HERMES $\nu$
and $Q^2$ kinematic region the expression (9) from \cite{kop4} turn in
formula (7) of
our work. Parameter $C^{\hspace{1mm}^{\prime}}$ which obtained in this
manner from dipole cross section is weakly changing function over $\nu$   
and $Q^2$, with the middle value of $C^{\hspace{1mm}^{\prime}}$ $\sim$ 1
mb$\cdot$Ge$V^2$. However, taking into account, that in energy
range $W^2$ = 4 $\div$ 50 Ge$V^2$ the value of ${\sigma}_{{\pi}N}^{tot}$ 
from \cite{kop4} seems to be underestimated, we are using
${\sigma}_{{\pi}N}^{tot}$ = 25 mb, and obtained for
$C^{\hspace{1mm}^{\prime}}$ numerical value close to C.\\
\hspace*{1em}In fact to describe the attenuations measured experimentally
as function of $\nu$, $Q^2$, or $z_h$, we are integrating the expression
(2) ower two of three mentioned above variables.
\hspace*{1em}The  calculations were 
performed using the following values for total cross section of hadron
\textbf{h} with
internuclear nucleon (we suppose that cross section for the
interaction of hadron with proton and neutron are equal):\\

\hspace*{1em}$\sigma_{\pi^+}$ = $\sigma_{\pi^-}$ = 25 mb; $\sigma_{k^+}$ =
17 mb; $\sigma_{k^-}$ = 23 mb.\\

\hspace*{1em}Some authors are using the inelastic cross section instead of
total ones \cite{bil3}. Unfortuntaly, the available experimental
information does not allow to make a choice between these two
posibilities.\\
\hspace*{1em}In present work the value of 1 GeV/fm is used for
the string tension.\\
\hspace*{1em}Also for all calculations the following kinematical conditions 
were used:\\
\hspace*{1em}$P_h$ $>$ 0.5 GeV,\hspace*{1em} $Q^2$ $>$ 1 Ge$V^2$, 
\hspace*{1em}$\nu$ $>$ 4 GeV,\hspace*{1em}$W^2$ $>$ 4 
Ge$V^2$,\hspace*{1em}y $<$ 0.85, 
\hspace*{1em}$z_h$ $>$ 0.1\\
where $P_h$ is the hadron momentum, $W^2$ is the hadron invariant mass square, 
y=$\nu$/$E_0$, $E_0$ is the incident lepton energy ($E_0$ = 27.5 GeV).\\
\hspace*{1em}With the presense of more data it will be possible to make a
global fit to determine the 
optimal values of string tension, the parameter
\textbf{C} in quark-nucleon cross section as well as to precise the value
of $\sigma_{\pi^{\pm}}$ and $\sigma_{k^{\pm}}$ cross-sections.\\

\bf
\Large
\hspace*{1em}2.\hspace*{1em}Results and Discussion\\

\normalsize
\hspace*{1em}The obtained results for the nuclear attenuation 
of charged pions and kaons ($R_{A}^{\pi(k)}$) calculated for four 
different targets(N, Ne, Kr, Xe) are 
presented on the figures 1-6. It is clearly seen that in case of pions 
attenuation as a function of $\nu$ there is no difference for opposite 
charges of pions. In 
case of kaons this difference becames noticeable at low values of 
$\nu$ ($\nu$ $<$ 12 GeV). Also one can see that the attenuation for 
$\pi$(K$)^-$ in case of 
symmetric targets is larger than for $\pi$(K$)^+$. While in case of non 
symmetric 
targets the situation is vice versa. For $z_h$ dependence one can 
note that 
the difference between opposite charged hadrons is increased for pions, 
particularly 
for heavy targets. Because of mainly HERMES Collaboration \cite{air,val} 
has produced 
the data with different charged pion's  attenuations  during 
last years, the calculated curves were tuned to the HERMES 
kinematics. The following additional cuts  were applied: $\nu$ $>$ 7 GeV, 
$z_h$ $>$ 0.2, 4.0 $<$ $P_h$ $<$ 13.5 and 4.0 $<$ $P_h$ $<$ 15.0 GeV
for N and Kr, respectively. One can see good enough
agreement for the predicted 
$\nu$-dependence of ${\pi}^{\pm}$, 
attenuations in case of N and Kr (Fig. 7a, c). For $z_h$ dependence only
sum of ${\pi}^+$ and 
${\pi}^-$ attenuations has been published and except one point there is 
also good agreement for calculated values and data (Fig. 7b).\\  
In conclusion, we have to note that developed approach can provide much 
better description of data after the fitting procedure based on all
existing data for 
different type of hadrons and targets. The parton distributions 
used for this approach \cite{glu} could be modified using the last results 
on 
parametrization for special case of nuclear parton distributions. As it 
was discussed above in the valence region the quark is knocked 
out from the nucleon, so the color string is streched between this quark 
and nucleon remnant, and the final hadrons are produced from the string. 
In the  "sea"  region mainly the "sea" 
(anti)quark is knocked out from the nucleon and two
strings are created. One of them with the knocked out (anti)quark takes away
the main part of virtual photon energy passing the next string an energy 
$\Delta$E which is enough to produce 1-2 hadrons or one resonance. In the 
framework of Regge model such production of hadrons from the big and 
little strings is called "Undeveloped Pomeron".  This mechanism could lead
to some changes of the attenuation values (particularly 
at low $\nu$), as well as to relative changes for
different 
hadrons and charges for the same hadron. As it is followed from 
the Regge model the $\Delta$E value is estimated to be 1 - 3 GeV and this 
value also should be fitted. The preliminary 
estimations of possible influence of the mentioned above scheme with 
$\Delta$E $\sim$ 1 - 3 GeV showed that they expected changes in   obtained
results will be not essential.\\
In generally we have to note that the attenuation values for different 
types of hadrons as well as for the oppositely charged 
mesons of the same kind are 
very close. For the most heavy target $^{131}$Xe the difference between 
the 
charged mesons is less than 5 $\%$, and for different mesons is less than 
10 $\%$.
In the same time one can see  strong enough $\nu$ dependence for 
attenuations within the HERMES kinematics range.\\

\hspace*{1em}Acknowledgments\\

\hspace*{1em}We would like to thank  our colleagues from HERMES
collaboration for useful discussions.

\newpage
\begin{appendix}
\bf
\Large
\hspace*{1em}Appendix\hspace*{1em}A: Functions 
$C^{h}_{fi}$(A,$x_{Bj}$)\\

\normalsize

Notations:
i = 1,2,3 means the ordinal number of three hadrons produced last in 
string(and the first number corresponds the last one)\\

\hspace*{1em}a) $\pi^+$(u\={d})\\
\begin{eqnarray}
C_{f1}^{\pi^+}=
\{\frac{Z}{A}[\frac{4}{9}(u_{v}(x_{Bj},Q^{2})+u_{s}(x_{Bj},Q^{2}))
+\frac{1}{9}{\bar d}_{s}(x_{Bj},Q^2)]\\ \nonumber
+\frac{N}{A}[\frac{4}{9}(d_{v}(x_{Bj},Q^{2})+d_{s}(x_{Bj},Q^{2}))
+\frac{1}{9}{\bar u}_s(x_{Bj},Q^2)]\}\gamma_{q}
\end{eqnarray}
\hspace*{1em}where \textbf{A} - is atomic number; \textbf{Z} and 
\textbf{N} - numbers of protons and neutrons in nuclei; $\gamma_{q}$ - is 
the probability of light
\textbf{q}$\textbf{\={q}}$ pairs 
(\textbf{u}$\textbf{\={u}}$,\textbf{d}$\textbf{\={d}}$)production in 
color field 
\cite{sjo}; $u_v$($x_{Bj}$,$Q^2$), 
$d_v$($x_{Bj}$,$Q^2$), $u_s$($x_{Bj}$,$Q^2$) etc. - parton distribution 
functions in proton. We used parametrization of \textbf{NLO({\={M}}S)} 
parton distributions from \cite{glu}.\footnote{In further (for brevity) we 
omit arguments of parton distribution functions}\\
\begin{eqnarray}
C_{f2}^{\pi^{+}}=
{\{}{\frac{Z}{A}}[{\frac{4}{9}}(u_{v}+u_{s}+{{\bar u}_{s}}) 
+ {\frac{1}{9}}(d_{v}+d_{s}+{{\bar d}_{s}}+s_{s}+{{\bar s}_{s}})]\\ \nonumber
+ \frac{N}{A}[{\frac{4}{9}}(d_{v}+d_{s}+{\bar d}_{s})
+ 
{\frac{1}{9}}(u_{v}+u_{s}+{\bar u}_{s}+s_{s}+{\bar s}_{s})]{\}}{\gamma}_{q}^{2}
\end{eqnarray}
\hspace*{1em}$C^{{\pi}^{+}}_{f3}$ = $C^{{\pi}^{+}}_{f2}$\\

\hspace*{1em}b) $\pi^-$(d\={u})\\ 
\begin{eqnarray}
C_{f1}^{\pi^{-}}=
{\{}\frac{Z}{A}[\frac{1}{9}(d_{v}+d_{s}) 
+ {\frac{4}{9}}{\bar u}_{s}]
+ \frac{N}{A}[
{\frac{1}{9}}(u_{v}+u_{s})+\frac{4}{9}{\bar d}_{s})]{\}}{\gamma}_{q}
\end{eqnarray}
\hspace*{1em}$C^{{\pi}^{-}}_{f2}$ and $C^{{\pi}^{-}}_{f3}$ the same as in 
previous case.\\

\hspace*{1em}c) $k^+$(u\={s})\\
\begin{eqnarray}
C_{f1}^{k^{+}}=
\frac{Z}{A}[{\frac{4}{9}}(u_{v}+u_{s}){\gamma}_{s} +
{\frac{1}{9}}{\bar s}_{s}{\gamma}_{q}]
+ {\frac{N}{A}}[\frac{4}{9}(d_{v}+d_{s}){\gamma}_{s}+
{\frac{1}{9}}{\bar s}_{s}{\gamma}_{q}]
\end{eqnarray}
\hspace*{1em}$C^{{k}^{+}}_{f2}$ = $C^{{k}^{+}}_{f3}$ =
$C^{{\pi}^{+}}_{f2}$$\frac{{\gamma}_{s}}{{\gamma}_{q}}$\\
\hspace*{1em}where ${\gamma}_{s}$ - is the probability of 
\textbf{s}$\textbf{\={s}}$ pairs production in 
color field  \cite{sjo}.\\

\hspace*{1em}d) $k^-$(s\={u})\\
\begin{eqnarray}
C_{f1}^{k^{-}}=
{\frac{Z}{A}}[{\frac{1}{9}}{s_{s}}{\gamma}_{q} +
{\frac{4}{9}}{\bar u}_{s}{\gamma}_{s}]
+ {\frac{N}{A}}[{\frac{1}{9}}{s_{s}}{\gamma}_{q}+
{\frac{4}{9}}{\bar d}_{s}{\gamma}_{s}]
\end{eqnarray}  
\hspace*{1em}$C^{{k}^{-}}_{f2}$ = $C^{{k}^{-}}_{f3}$ =
$C^{{k}^{+}}_{f2}$\\
\hspace*{1em}For the all described here hadrons: $C^{h}_{fi}$ = 
$C^{h}_{f2}$, where \textbf{i} = 3, 4, 5  \\

\bf
\Large
\hspace*{1em}Appendix\hspace*{1em}B: The distribution of constituent\\ 
\hspace*{1em}formation lengths in Field-Feynman fragmentation scheme\\

\normalsize
\hspace*{1em}$D_{i}^{c}$(x,$l_c$) - is the distribution function of 
constituent formation lengths $l_c$, for i-th hadron in 
string 
carry away parth of energy \textbf{x} of virtual photon. To obtain 
these functions in \cite{bil3} it was used the function f(x), which have 
the meaning 
of probability, that the first hierarchy (rank - 1) primary meson leaves 
the fraction of momentum \textbf{x} to the remaining cascade. 
We are using function f(x) obtained in \cite{fie}: f(x) = 1 - a +
3a$x^2$\\
\hspace*{1em} As it is well known the $\pi$ and K mesons could be produced 
in 
direct way 
and also as a result of resonances decay. In this work only the direct 
production is taking in account. In case of pions it was estimated the 
contribution coming from the $\rho$ meson decay and it was shown that 
the 
second mechanism (via resonances) did not change the values of calculated 
attenuations.\\
\hspace*{1em}As noticed above, we are limited only with three hadrons on 
the fast 
end of string. For these hadrons the distribution functions are given as\\
\begin{eqnarray}
D_{1}^{c}(x,{l_c})=
f(1-x)\delta[{l_c}-(1-x)L]
\end{eqnarray}  
\begin{eqnarray}
D_{2}^{c}(x,{l_c})=
{\frac{1}{x+{\frac{l_c}{L}}}}f({\frac{\frac{l_c}{L}}{x+{\frac{l_c}{L}}}})
{\frac{1}{L}}f(x+{\frac{l_c}{L}})
\end{eqnarray}  
\begin{eqnarray}
D_{3}^{c}(x,{l_c})=
{\frac{1}{x+{\frac{l_c}{L}}}}f({\frac{\frac{l_c}{L}}{x+{\frac{l_c}{L}}}})
{\frac{1}{L}}
{\int_{u_1min}^{l}}{\frac{du_1}{u_1}}f(u_1)
f({\frac{{l_c}+xL}{L{u_1}}})
\end{eqnarray}

\hspace*{1em}where $u_{1min}$ = $\frac{{l_c}+xL}{L}$; L - length of string 
(L = $\nu$/k).\\
\hspace*{1em} The general formula for $D_{i}^{c}$(x,$l_c$) is given as\\
\begin{eqnarray}
D_{i}^{c}(x,{l_c})=
\frac{1}{x+{\frac{l_c}{L}}}f(\frac{\frac{l_c}{L}}{x+{\frac{l_c}{L}}})
\frac{1}{L}
\int_{u_{1min}}^{l}\frac{du_1}{u_1}f(u_1)...
\int_{u_{i-{2min}}}^{l}\frac{du_{i-2}}{u_{i-2}}f(u_{i-2})
f(\frac{{l_c}+xL}{L{u_1}...{u_{i-2}}})
\end{eqnarray}

\hspace*{1em}where \textbf{i} = 3, 4, 5, . . ., and $u_{jmin}$=
${\frac{{l_c}+xL}{L{u_1}{u_2}...{u_{j-1}}}}$\\

\bf
\Large
\hspace*{1em}Appendix\hspace*{1em}C: The functions $F_i$(x)\\

\normalsize
\hspace*{1em}$F_i$(x) - is the probability, that the \textbf{i} - th 
hadron in string carry away the fraction of momentum \textbf{x}.\\
\hspace*{1em}The general formula for $F_{i}$(x) is given as\\
\begin{eqnarray}
F_i(x)=
{\int_{x}^{l}}{\frac{d\eta_1}{\eta_1}}f(\eta_1)
{\int_{x/{\eta_1}}^{l}}{\frac{d\eta_2}{\eta_2}}f(\eta_2)...
{\int_{x/{\eta_1}...{\eta_{i-2}}}^{l}}{\frac{d\eta_{i-1}}{\eta_{i-1}}}
f(\eta_{i-1})f({1-{\frac{x}{\eta_1\eta_2...\eta_{i-1}}}})
\end{eqnarray}  
\hspace*{1em} For cases of \textbf{i} = 1, 2, 3\\
\begin{eqnarray}
F_1(x)= f(1 - x)
\end{eqnarray}
\hspace*{1em}where f(x) was definded in Appendix B
\begin{eqnarray}
F_2(x)= {\int_{x}^{1}}{\frac{d\eta}{\eta}}f(\eta)f(1 - {\frac{x}{\eta}})
\end{eqnarray}
\begin{eqnarray}
F_3(x)= {\int_{x}^{1}}{\frac{d\eta}{\eta}}f(\eta)
{\int_{x/{\eta}}^{1}}{\frac{d\eta_1}{\eta_1}}f(\eta_1)f(1 - 
{\frac{x}{\eta{\eta_1}}})
\end{eqnarray}
\end{appendix}

\newpage

\begin{figure}[htb]
\unitlength 1mm
\begin{picture}(170,180)
\put(6,0){
\epsfxsize=13cm
\epsfysize=15cm
\epsfbox{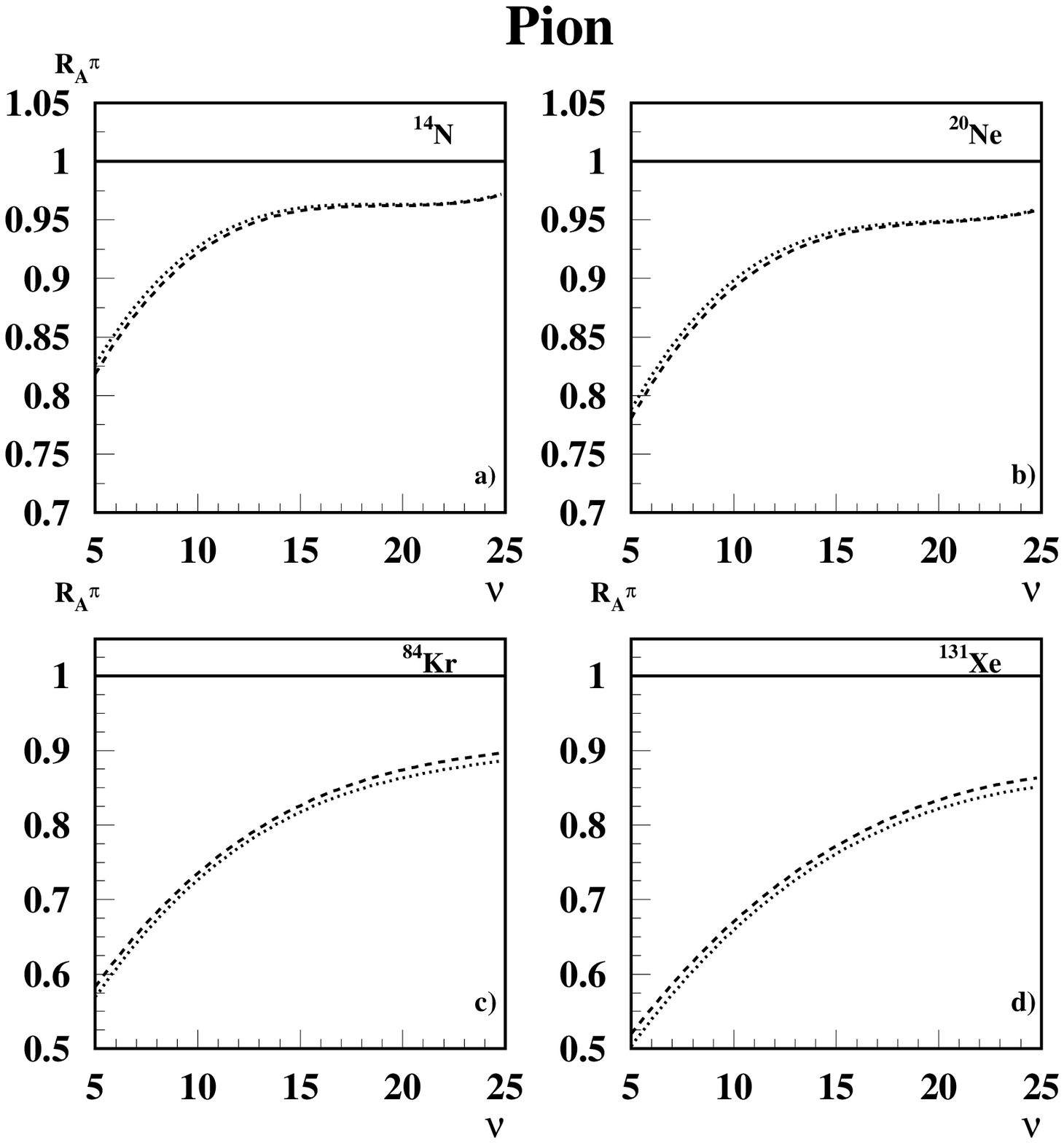}
}
\end{picture}
\caption{\protect\it
The $\nu$-dependence of the ratio $R_{A}^{\pi}$ for the different 
targets: a)$^{14}$N; b)$^{20}$Ne; c)$^{84}$Kr; d)$^{131}$Xe. The dashed 
curves
correspond to the case ${\pi}^-$ and dotted - ${\pi}^+$ mesons }  
\label{tp}
\end{figure}

\begin{figure}[htb]
\unitlength 1mm
\begin{picture}(180,170)
\put(6,0){ 
\epsfxsize=13cm
\epsfysize=16cm 
\epsfbox{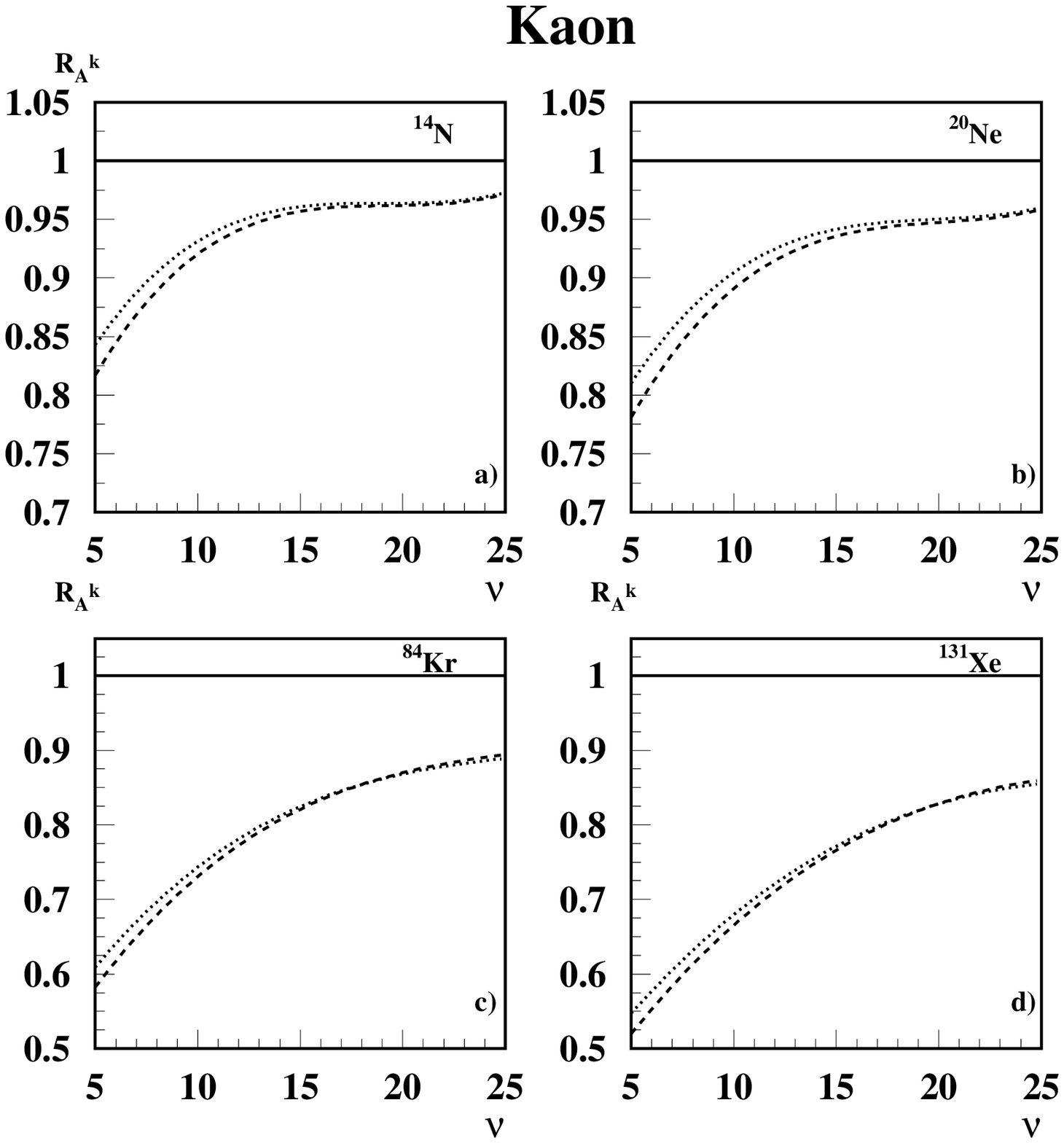}
}
\end{picture}  
\caption{\protect\it
 The same as in Fig. 1, but calculated for the charged kaons }
\label{tp}
\end{figure}
\begin{figure}[htb]
\unitlength 1mm
\begin{picture}(180,200)
\put(6,0){ 
\epsfxsize=13cm
\epsfysize=14cm 
\epsfbox{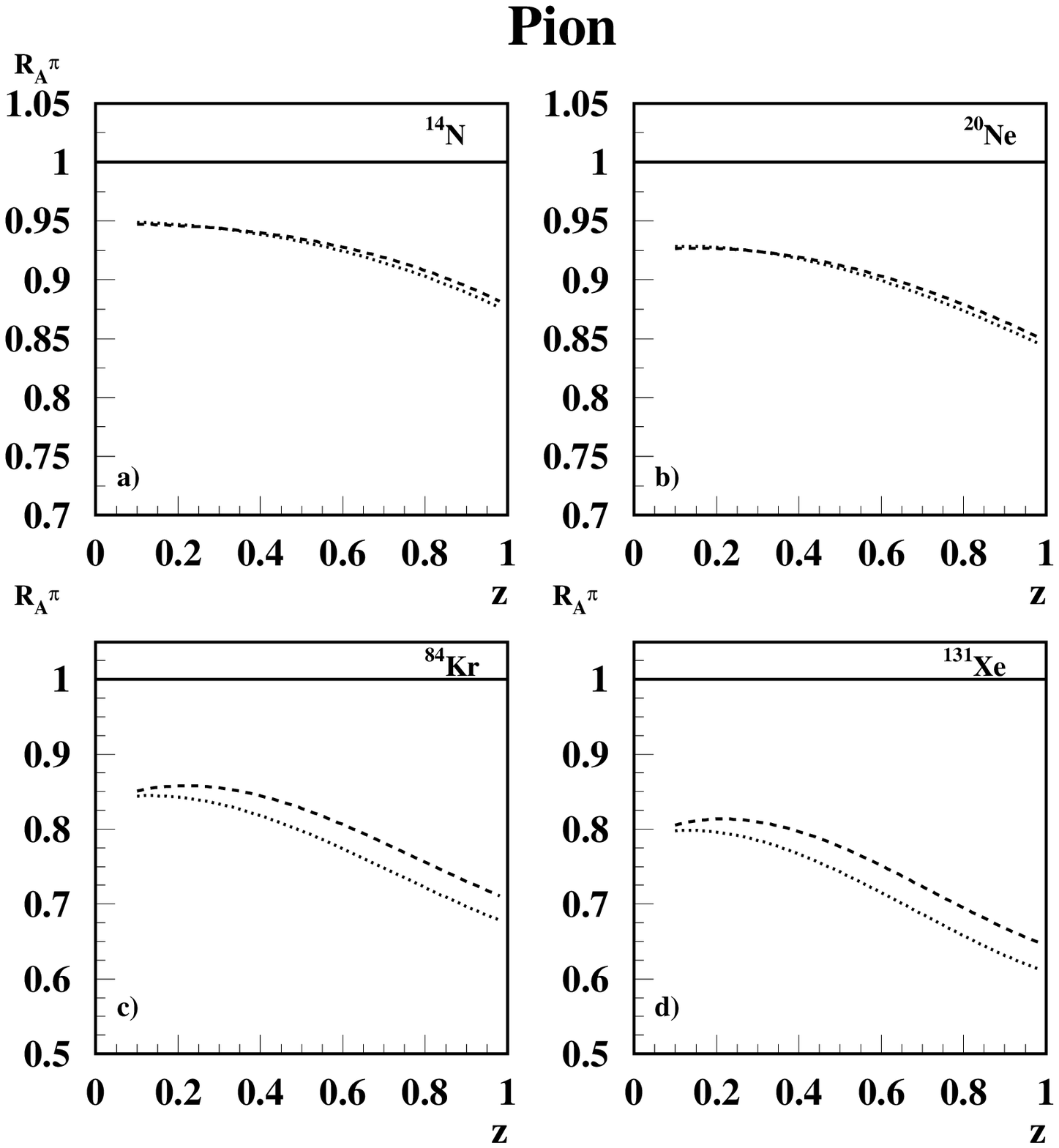}
}
\end{picture}  
\caption{\protect\it
 The $z_h$ - dependence of the ratio $R_{A}^{\pi}$ for the different
targets: a)$^{14}$N; b)$^{20}$Ne; c)$^{84}$Kr; d)$^{131}$Xe. The dashed
curves
correspond to the case ${\pi}^-$ and dotted - ${\pi}^+$ mesons }
\label{tp}
\end{figure}
\begin{figure}[htb]
\unitlength 1mm
\begin{picture}(180,200)
\put(6,0){ 
\epsfxsize=13cm
\epsfysize=14cm 
\epsfbox{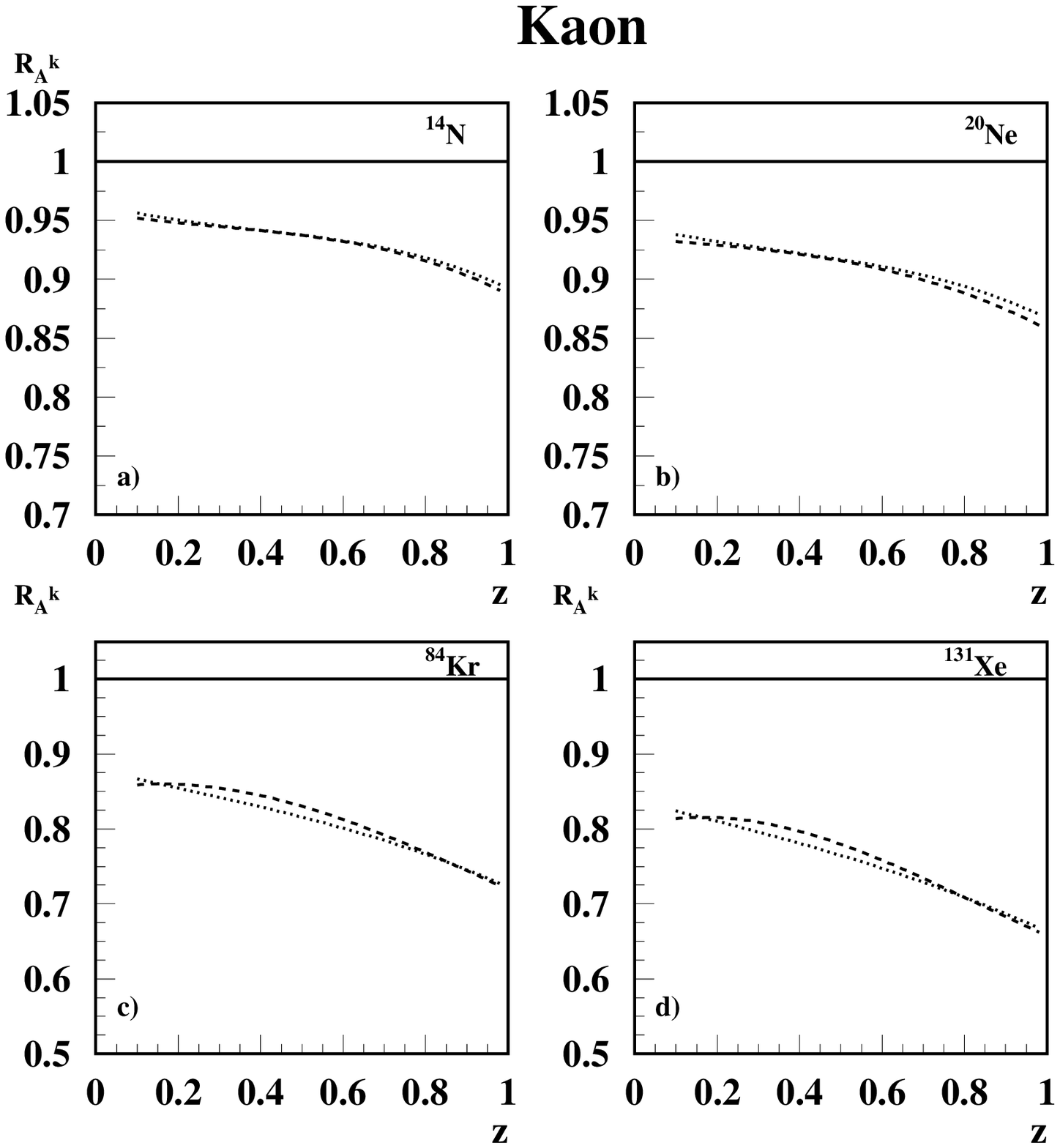}
}
\end{picture}  
\caption{\protect\it
 The same as in Fig. 3, but calculated for the charged kaons }
\label{tp}
\end{figure}
\begin{figure}[htb]
\unitlength 1mm
\begin{picture}(180,200)
\put(6,0){ 
\epsfxsize=13cm
\epsfysize=14cm 
\epsfbox{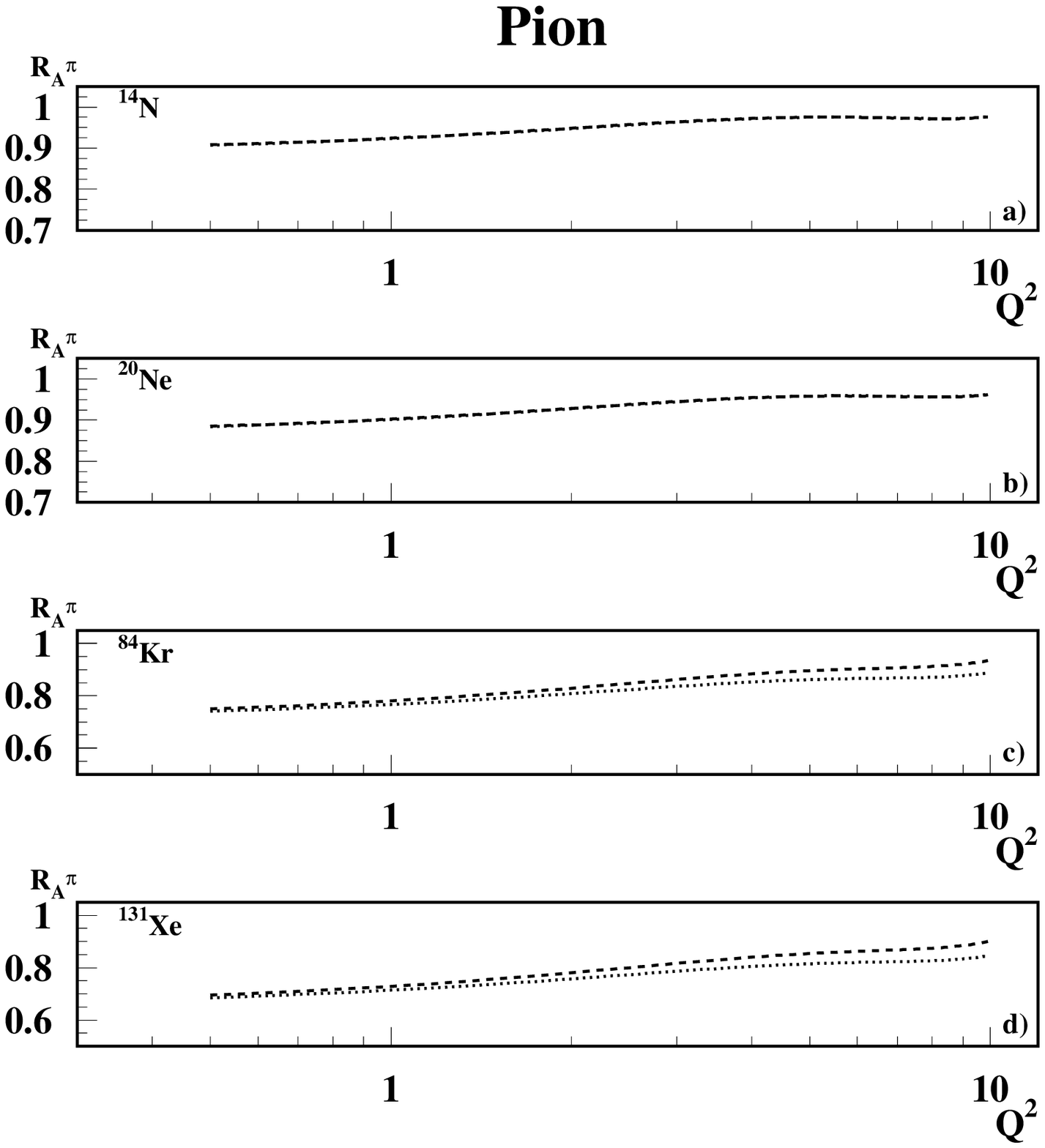}
}
\end{picture}  
\caption{\protect\it
 The $Q^2$ - dependence of the ratio $R_{A}^{\pi}$ for the different
targets: a)$^{14}$N; b)$^{20}$Ne; c)$^{84}$Kr; d)$^{131}$Xe. The dashed
curves
correspond to the case ${\pi}^-$ and dotted - ${\pi}^+$ mesons }
\label{tp}
\end{figure}
\begin{figure}[htb]
\unitlength 1mm
\begin{picture}(180,200)
\put(6,0){ 
\epsfxsize=13cm
\epsfysize=14cm 
\epsfbox{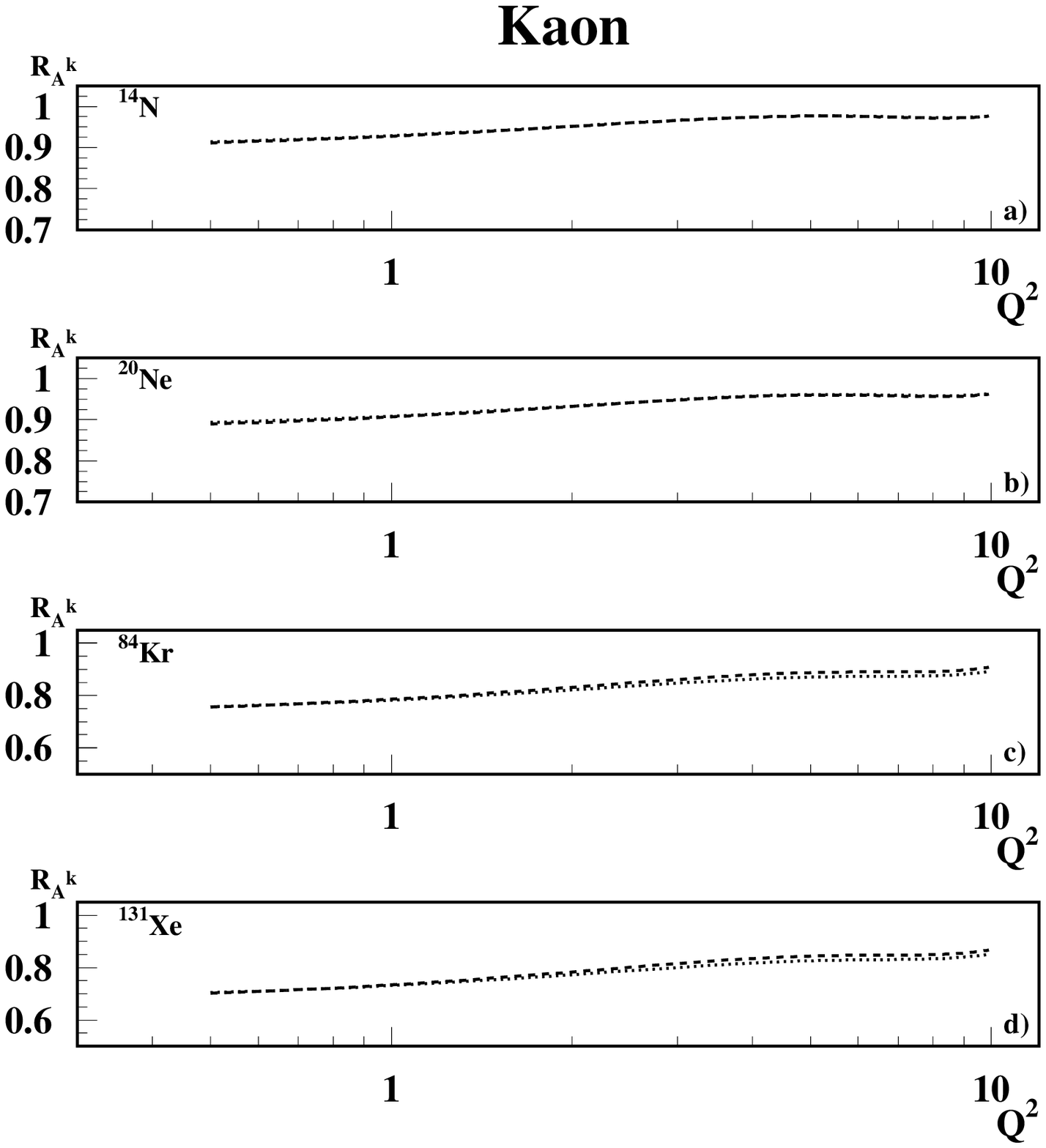}
}
\end{picture}  
\caption{\protect\it
 The same as in Fig. 5, but calculated for the charged kaons }
\label{tp}
\end{figure}
\begin{figure}[htb]
\unitlength 1mm
\begin{picture}(200,180)
\put(5,0){
\epsfxsize=14cm
\epsfysize=16cm
\epsfbox{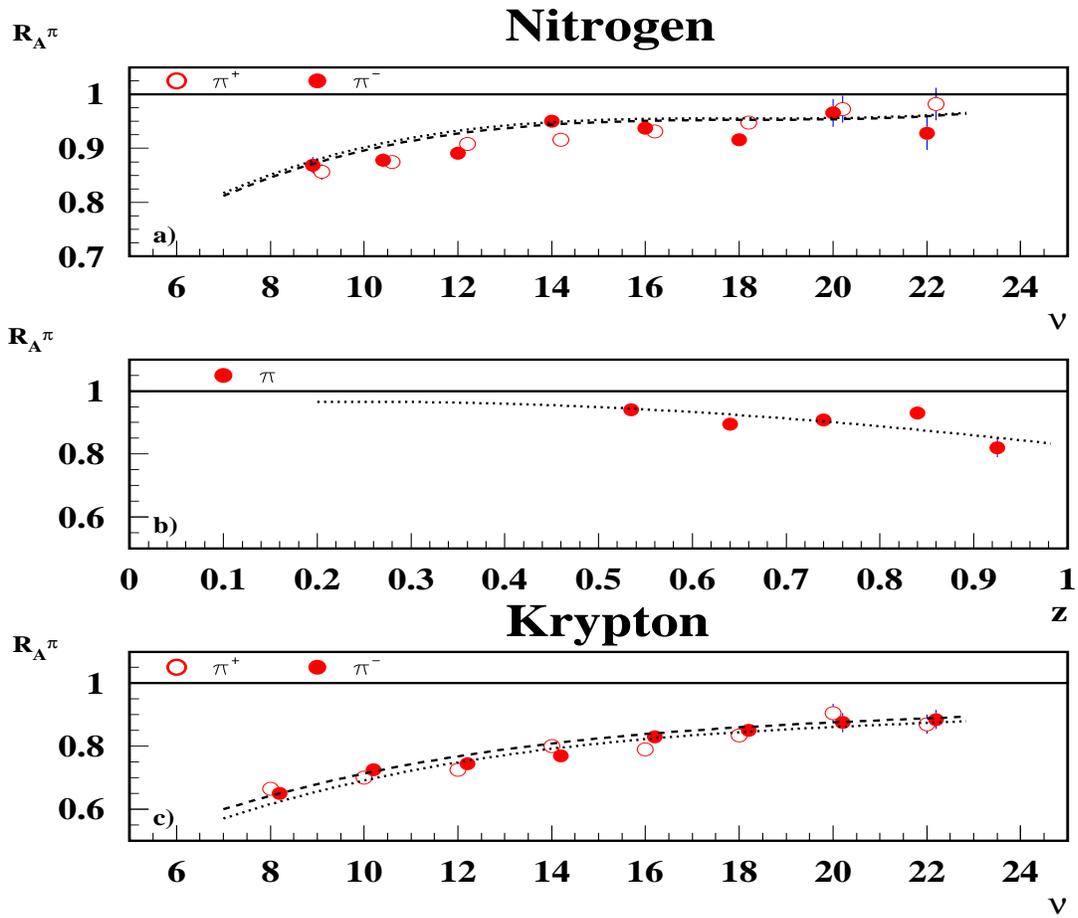}
}
\end{picture}
\caption{\protect\it
The comparison of predicted values for attenuation with HERMES data:
a)$\pi^{\pm}$ vs $\nu$ for $^{14}$N \cite{air}; b) $\pi^+$ + $\pi^-$ vs
$z_h$ 
for $^{14}$N \cite{air}; c)$\pi^{\pm}$ vs $\nu$ for $^{85}$Kr
\cite{val}(preliminary results)}    
\label{tp}
\end{figure}
\end{document}